\documentclass[%
nofootinbib,
reprint,
amsmath,amssymb,
aps,10pt
]{revtex4-2}

\usepackage{graphicx}
\usepackage{dcolumn}
\usepackage{bm}

\usepackage{xcolor}

\newcommand{\tr}{\operatorname{tr}}
\begin{document}
	\preprint{PU-}
	
	\title{Hidden Wave Function of Twisted Bilayer Graphene: Flat Band as a Landau Level}
	
	\author{Fedor K. Popov}
	\altaffiliation{Princeton University, Physics Department; Moscow Institute for Physics and Technology}
	\email{fedor.popov@phystech.edu}
	\author{Alexey Milekhin}
	\altaffiliation{UC Santa Barbara, Physics Department}
	\email{milekhin@ucsb.edu}
	\date{\today}
	
	\begin{abstract}
	We study zero energy states of the chirally symmetric continuum model (CS-CM) of the twisted bilayer graphene.
	The zero energy state obeys Dirac equation on a torus in the external non-abelian magnetic field. These zero energy states could form a flat band --- a band where the energy is constant across the Brillouin zone. 	We prove that the existence of the flat band implies that the wave-function of any state from the flat band has a zero and vice verse. 
	We found a hidden flat band of unphysical states in the CS-CM model that has a pole instead of a zero.
	Our main result is that in the basis of the flat band and hidden wave functions the flat band could be interpreted  as  Landau level in the external magnetic field. From that interpretation we show the existence of extra flat bands in the magnetic field.
	\end{abstract}
	\maketitle
	{\it Introduction.}
	Twisted bilayer graphene (TBG) has recently drawn a lot of attention
from the physics community due to its interesting properties and applications
\cite{Cao2018a} - \cite{WolfLado}. One of the most prominent features is the
recent discovery of correlated insulators and superconductivity, that are
observed in a narrow range of twist angles near $\theta=1.05^\circ$ , which is
usually referred to as the magic angle.  At this angle the system develops a
nearly flat band near  charge neutrality.  Recently the flat band was
explored analytically in a chiral model of TBG that neglects the
hoppings within the same sublattices of different TBG sheets  \cite{Grisha}.
In this paper we continue the exploration of the mathematical structures of the
flat band and demonstrate the connection with the vector bundles over the
Riemann surfaces of higher genus that could provide some deeper understanding of the physics behind CS-CM model of twisted bilayer graphene. From physical standpoint this will allow
us to study the behavior of flat bands in external magnetic field.

	TBG consists of a two graphene sheets placed on top of each other at small angle $\theta \ll 1$ that form a long-period pattern (moire pattern). One can estimate that the period of the resulting super-lattice is of the order $L(\theta) \sim \frac{a}{\theta} \gg a$, where $a$ is the graphene lattice constant. That allows us to consider a continuum model for the Hamiltonian instead of a lattice one. This approach was used  by Bistrizer and MacDonald \cite{Bistritzer2010,Bistritzer2011} and  by Lopes dos Santos \cite{Castro-Neto2012}. Thus we can write an effective Hamiltonian for this model \cite{Grisha} as
	\begin{gather*}
		H_0 = \begin{pmatrix}
			i v_0 \vec{\sigma}_{\theta/2} \vec{\nabla} & T(r) \\
			T^\dagger(r) &  i v_0 \vec{\sigma}_{-\theta/2} \vec{\nabla}
		\end{pmatrix}, \, T(r) = \begin{pmatrix}
		t_{aa}(r) & t_{ab}(r)\\
		t_{ba}(r) & t_{bb}(r)
	\end{pmatrix}
	\end{gather*}
	where we already used the fact that the superlattice is much bigger that the interatomic lattice of separate sheets of graphene.  Therefore we can use Dirac equation to describe excitations in each individual graphene sheet. The off-diagonal term $T$ is responsible for hopping between sheets of the TBG and sublattices $a$ and $b$ of the individual graphene sheets. The $H_0$ acts on the four dimensional wave function $\Psi = \left(\psi_{a1},\psi_{b1},\psi_{a2},\psi_{b2}\right)^T$, where the second index relates to the individual graphene sheets of the TBG and the first index relates to the sublattice of the given graphene sheet.

    The numerical study of that model confirmed the existence of the flat band at magic angle $\theta^*_1 \approx 1.05^{\circ}$.  	The modification studied here  neglects the coupling between the sublattices of the graphene $t_{aa}=t_{bb}=0$. In this case system acquires an additional chiral symmetry and usually referred to as \textit{chirally symmetric continuum model} (CS-CM). After an appropriate change of basis the Hamiltonian of CS-CM can be casted in the following form
	\begin{gather}
	H = U H_0 U^{-1} = \begin{pmatrix}
	0 & \mathcal{D}\\
	\mathcal{D}^* & 0,
	\end{pmatrix},\quad {\rm where} \notag\\
    \mathcal{D} = \begin{pmatrix}
	2 i \bar{\partial} + W(\vec{r}) &  V(\vec{r})\\
	U(\vec{r}) & 2i \bar{\partial} - W(\vec{r})
	\end{pmatrix}
\label{tbgHam}
	\end{gather}
	where $\vec{r}$ is the vector in 2d graphene sheet, $\bar{\partial} = \frac12\left(\partial_x - i \partial_y\right)$ is anti-holomorphic derivative along the sheet and $V(r),U(r)$ are the hopping potentials between the two sheets of the TBG that could be expressed linearly through $t_{ab}(r)$ and $t_{ba}(r)$. This Hamiltonian acts on the rotated wave functions $\Psi_U = U \Psi=\left(\phi,\psi\right)^T=\left(\phi_1,\phi_2,\psi_1,\psi_2\right)^T$.

The spectrum of the model is governed by the following eigenvalue problem
	\begin{gather}
	H \Psi_U = E \Psi_U \Leftrightarrow \left\{ \begin{matrix}
		\mathcal{D} \psi = E \phi\\
		\mathcal{D}^* \phi = E \psi
	\end{matrix} \right .
	\end{gather}	
	Since we are interested in the existence of a flat band near charge neutrality these equations simplify and we should study only the following equation
	\begin{gather}
	-\frac{i}{2}\mathcal{D} \psi = \left(\bar{\partial} + \bar{A}\right) \psi=0, \quad \mathcal{D}^* \phi = 0\label{fundeq}  \\ \text{where }\bar{A} = -\frac{i}{2}\begin{pmatrix}
	-W(\vec{r}) & V(\vec{r}) \\
	U(\vec{r}) & W(\vec{r})
	\end{pmatrix}, \notag
	\end{gather}
	due to the enhanced chiral symmetry the equations on $\psi$ and $\phi$
are decoupled, which allows for a deeper analytical investigation of the
properties of the CS-CM. These equations could be interpreted as a Dirac
equation in a non-abelian magnetic field $\bar{A} \in \mathbf{su}(2)$ on a
Riemann surface \cite{Guinea2012}. We will use this interpretation to bolster
our intuition and draw interesting conclusions. Since we study a periodic
system we should impose Bloch boundary\footnote{Please note that in \cite{Grisha} the authors introduced additional matrix twist to the boundary conditions. One can show that the results of our paper are not affected by this change, therefore for simplicity of the argument we will consider only boundary conditions of the form \eqref{bndc}.}
conditions
	\begin{gather} \label{bndc}
	\psi_{\vec{k}}(\vec{r} + \vec{a}_{1,2}) = e^{i \vec{k} \vec{a}_{1,2}} \psi_{\vec{k}}(\vec{r}),
	\end{gather}
$\vec{a}_{1,2}$ are the periods of the moire super-lattice and the vector $\vec{k}$ defines the location in the moire Brillouin zone (MBZ). 
If the solution exists for any point $k$ in MBZ zone then the system has a \underline{\it flat band}.
One can show that such potentials exists \cite{Grisha}. But for a general chosen potential $\bar{A}$, the system of equations \eqref{fundeq} and \eqref{bndc} has a smooth finite solution {\it only} at finite numbers of points $k$ in MBZ.

\textit{The purpose of this paper} is to consider a \underline{generic potential} $\bar{A}$ in the equation \eqref{fundeq} and get general properties independent of the concrete form of $\bar{A}$, that could shed the light on the physics behind the CS-CM model. Our main result is that once a system posses flat band we can separate TBG into a system of two individual sheets with positive and negative effective magnetic fields, that supports Landau levels of different chirality \cite{PhysRevB.31.2529}. The negative magnetic field could be canceled by an external magnetic field, resulting in additional flat bands. Therefore, the number of flat bands increases in the presence of the magnetic field. \textit{This is the main physical result of this paper.}

\textit{The paper is organized as follows.}
First, we will build an integral of motion of the equation \eqref{fundeq} which, as was shown in \cite{Grisha}, is related to the Fermi velocity. Hence we will refer to it as a Fermi integral of motion $I_F$. We prove that the flat band appears if and only if this invariant is equal to zero, $I_F=0$.
Then we demonstrate that the system of equations \eqref{fundeq} and \eqref{bndc} admits an additional solution, which is singular and therefore unphysical. 
This second solution will allow us to rewrite the system of equations in the form of two Dirac equations on a torus with effective magnetic fields. This shows the direct connection of flat band to the Landau levels. 
Finally, we will show that introducing an external magnetic field can make the second solution non-singular. Hence this leads to additional flat bands which, in principle, could be seen on the experiment. The mathematical details are delegated to the Appendix.

{\it Fermi integral, zeros of wave functions and the flat band.}	
	For simplicity, let us consider  equation \eqref{fundeq} alone without taking into account the boundary conditions \eqref{bndc}:
	\begin{gather} \label{tbgEq}
	\mathcal{D}\psi = \left(\bar{\partial} + \bar{A}\right)\psi=0,\, \psi = \left(\psi_1,\psi_2\right)^T \in \mathbb{C}^2, \,\tr \bar{A}=0
	\end{gather}
	We start from studying the properties of the vector-valued function $\psi$ that satisfy equation $\mathcal{D}\psi=0$. Such equations have been broadly studied in some fields of mathematics. Hence, to simplify the further computation and exploit the results we adopt some mathematical terminology. We assume that our TBG is separated into geometric domains $U_\alpha$ such that when we jump from one domain to another we should appropriately change the vector-valued function $\psi$: $$\begin{pmatrix}\psi_{1\alpha} \\ \psi_{2\alpha} \end{pmatrix} = g_{\alpha\beta} \begin{pmatrix}\psi_{1\beta} \\ \psi_{2\beta} \end{pmatrix}.$$
	The collection $\psi=\left\{\psi_\alpha \right\}$ is said to be a section of a vector bundle $E$, which is a collection of domains $\left\{U_\alpha\right\}$ with translation functions $g_{\alpha\beta}$. If there is only one domain the bundle is said to be trivial. An example of non-trivial vector bundle is provided by separating the TBG into a set of fundamental domains by acting with translations $\vec{a}_{1,2}$. Translation functions in this case are boundary conditions \eqref{bndc}.

	From the mathematical point of view, the holomorphic equation $\bar{\partial}\psi=0$ is similar to the equation \eqref{tbgEq}: $\mathcal{D}\psi=0$. Then mathematicans say that if $\mathcal{D}\psi=0$ then the wave function $\psi$ is a {\it meromorphic} function and $\bar{A}$ is a holomorphic connection. If $\psi$ is also finite everywhere, we would call such a function {\it holomorphic}. The convenience of such terminology is that such $\psi$  share a lot of properties with usual holomorphic functions studied in the complex analysis. 

	From the physical point of view, any wave function must be finite. So we must assume that $\psi$ is also {\it a holomorphic} function in the above sense of vector bundle  \footnote{Strictly speaking, we can allow integrable singularities like $1/z^{1/4}$ but they are not consistent with eq. (\ref{tbgEq}) because we assume that $\bar{A}$ does not have poles.} $E$.

	Let us consider two finite solutions $\psi_1, \psi_2$ of the equation \eqref{tbgEq}. One can compute the Wronskian of these solutions
	\begin{gather}
	I_F(\psi_{1},\psi_{2}) = \det(\psi_{1}, \psi_{2}) = I_F(r), \quad \text{then}\notag\\
	\bar{\partial} I_F(r) = -\tr \bar{A} \cdot I_F(r), \quad \tr \bar{A} =0 \,\Rightarrow\, I_F(r) = I_F(z), \label{WronksConserv}
	\end{gather}
	where we have used the fact that $\bar{A} \in \mathbf{su}(2)$ and hence traceless. We come to the conclusion that the Wronskian $I_F(z)$ must be an analytic function. If $\psi_{1,2}$ are finite everywhere, $I_F(z)$ is holomorphic and therefore must be constant (because of the Liouville theorem) across the plane of TBG. Because of this property we can consider $I_F$ as an integral of motion of the equation \eqref{tbgEq}. This property could be generalized to other systems and will provide a necessary and sufficient condition for the existence of flat band in the system.

	\textit{From flat band to zero Wronskian $I_F=0$}.
	Here we prove that we cannot have a flat band unless $I_F=0$. Therefore applying negation, a flat band wavefunction has a zero and hence $I_F=0$.
	We will say in a minute which two solutions we need to pick up.
	For the application to TBG we should study the equation \eqref{fundeq} on a torus as it was explained in the previous section.
	Namely, we can consider TBG as a torus $\mathbb{C}/\Lambda$, where $\Lambda={m a_1 + n a_2, m,n \in \mathbb{Z}}, a_{1,2} = a^x_{1,2} +i a^y_{1,2}$. We must impose boundary conditions (\ref{bndc}) to glue the wave function as we shift along the lattice $\Lambda$. These boundary conditions are the gluing functions of the vector bundle over the torus. Without loss of generality we will set $a_1=1$ and $a_2=\tau$.

	We define $\mathbb{C}^2_K$ to be the vector bundle with boundary conditions(gluing functions) \eqref{bndc} with quasi-momentum $K$ . Again, equation \eqref{tbgEq} with the connection $\bar{A}$ defines a meromorphic section of this vector bundle.

		For the sake of argument, we assume that there are at least two points $K_1,K_2$ in the MBZ where the solution exists. We would like to stress that $K_{1,2}$ are different from special points $K,K'$ usually considered in the study of TBG, where due to discrete symmetry $C_3$ the band must have zero energy $E=0$. Of course, in the TBG the potential must respect the $C_3$ symmetry and therefore the Dirac points must exist at points $K,K'$. But in the general twisted bilayer material such symmetry could be absent and to make the discussion more general we make an assumption that at least two points $K_{1,2}$ where the gap closes. Therefore, to keep discussion general we just assume that due to some lucky choice of $\bar{A}$ in the equation \eqref{fundeq} a system has zero energy at some points $K_{1,2}$ of MBZ.
	
    The relation (\ref{WronksConserv}) still holds true as it is not sensitive to boundary conditions. If we have two holomorphic solutions $\psi_{K_{1,2}}$ at two different points of the Brillouin zone $K_{1,2}$  we can compute the Wronksian
	\begin{gather}
	I_{F,K_1+K_2}(z)= I_F(\psi_{K_1},\psi_{K_2}) = \det(\psi_{K_1}, \psi_{K_2}),
	\end{gather}
	but because the $I_F(z)$ is holomorphic and bounded in the complex plane of TBG (due to the periodicity conditions (\ref{bndc}) and the fact that
	$\psi_{K_1}, \psi_{K_2}$ are finite) we must conclude that $I_F(z)={\rm const}$. Moreover, using boundary conditions \eqref{bndc} we have
	\begin{gather}
	I_{F,K_1+K_2}(z+a_{1,2}) = I_{F,K_1+K_2}(z) e^{i (K_1+K_2) a_{1,2}},
	\end{gather}
	But if $K_1+K_2 \neq 0$ and $I_F(z)$ is constant, the boundary conditions are satisfied only if $I_F(z)=0$. Hence, there are only \underline{two possibilities}:
	\begin{enumerate}
		\item  $I_F(z)=0$ and $K_1,K_2$ are arbitrary.
		\item $I_F(z)\neq 0$ but $K_1=-K_2$.
	\end{enumerate}
	  
	  We start with the second possibility. We normalize the solutions such that $I_F=1$. Then we immediately get that $\psi_{K_1}, \psi_{-K_1}$ are nowhere zero, because otherwise the Wronksian would be equal to zero at points where $\psi_{\pm K_1} = \vec{0}$. 
	  Since $I_F(z)$ is non-zero, the solutions $\psi_{K_1}$ and $\psi_{-K_1}$ are linearly independent at each point of the TBG. If we consider now matrix $M=\left(\psi_{K_1}, \psi_{-K_1}\right)$ it satisfies the following equation
	  \begin{gather}\label{flatsl}
	  \left(\bar{\partial} + \bar{A}\right)M= 0,\quad \bar{A} = - \bar{\partial} M \cdot M^{-1},
	  \end{gather}
	  where we used the fact that if $\det M \neq 0$ the matrix $M$
is invertible. We would like to point out, that the equation \eqref{flatsl}
does not mean $\bar{A}$ is a pure gauge(and hence a flat connection), since $M \in SL(2,\mathbb{C})$ rather than
$SU(2)$ and therefore is not a genuine gauge transformations. 
  	  
	  Let us consider another solution $\psi$ of the eq. (\ref{tbgEq}). Since $\psi_{\pm K_1}$ are linearly independent we can always represent $\psi$ as a linear combination of these solutions:
	  \begin{gather*}
	  \psi = v_1(r) \psi_{K_1} + v_2(r) \psi_{-K_1},
	  \end{gather*}
  Applying the operator $D=\bar{\partial} + \bar{A}$ we get
  	  \begin{gather}
  	  D\psi = \bar{\partial}v_1\, \psi_{K_1} + \bar{\partial}v_2\, \psi_{-K_1} = 0.
  	  \end{gather}
    	Since $\psi_{\pm K_1}$ are linearly independent at each point of the torus $\mathbb{C}/\Lambda$ it follows that the  coefficients $v_i$ must be holomorphic $\bar{\partial} v_i =0$. The functions $\psi$ and $\psi_{\pm K_1}$ are finite and non-zero everywhere, hence $v_i$ are bounded. From the maximum principle for analytic functions on a complex plane, $v_i$ are constant. Therefore if we have an arbitrary solution of the equation \eqref{tbgEq} at  point $k$ of the Brillouin zone we must have
 	\begin{gather}
 		\psi_k = v_{k,K_1} \psi_{K_1} + v_{k,-K_1} \psi_{-K_1},\quad v_{k,\pm K_1} \in \mathbb{C},
 	\end{gather}
 		but it is easy to see that with any choice of numbers $v_{\pm K_1}$ we are not able to satisfy  boundary conditions (\ref{bndc}) in MBZ. Therefore we cannot have a flat band if $I_F \neq 0$. 
 		
 		\textit{From zero Wronskian to a flat band.} 
		Let us prove the converse. Namely, if $I_F=0$ for some points $K_1, K_2$ in
		 the MBZ, the system develops a flat band. In other words, the equation \eqref{tbgEq} has a solution at any point $k$ in the MBZ. 
		
		We start by noticing that since $I_F(z)=0$ and $\psi_{K_1,K_2}$ satisfy the equation \eqref{tbgEq}, then wave function $\psi_{K_1}$ has a zero. 
		Let us prove this statement by contradiction. Assume the opposite: that $\psi_{K_1}(r) \neq 0$ at any point of the torus, $\mathbb{C}/\Lambda$, or fundamental domain of TBG. Because torus is compact the minimum $\min\limits_{r \in \mathbb{C}/\Lambda}\left|\psi_{K_1}(r)\right| = m > 0$ is reachable. If  $I_F(z)=0$ the wave functions $\psi_{K_1,K_2}$ are proportional to each other:
	\begin{gather*}
	    \psi_{K_2}(r) = \gamma(r) \psi_{K_1}(r), \quad D \psi_{K_2} = \bar{\partial}\gamma(r) \psi_{K_1}(r)=0,
	\end{gather*}
	where $\gamma(r)$ is bounded as $\left|\gamma(r)\right| < \frac{\left|\psi_{K_2}(r)\right|}{m}$ and holomorphic $\bar{\partial} \gamma(r) =0$ .  Then the function $\gamma(r)=\gamma(z)$ must be constant by the maximum principle. However, this is impossible since $\psi_{K_1,K_2}$ satisfy different boundary conditions. Hence we must conclude that $\psi_K$ has at least one simple zero\footnote{Holomorphicity of $\gamma(z)$ concludes that the wave function has a simple zero $\psi_K(z) \sim f(\bar{z})\left(z - z_0\right) + \mathcal{O}(z-z_0)$ rather than some non-analytical behavior.}.

	We can now follow the procedure described in the paper \cite{Grisha} and construct solution at any point $k$ of the Brillouin zone. The specific potential studied in \cite{Grisha} had an extra property: $I_F \propto v_F$, and hence such solution implied the existence of a flat band. Our reasoning managed to generalize this condition to an arbitrary potential $\bar{A}$. 
	
{\it Hidden wave function.}	We can draw some additional conclusions from the existence of the holomorphic section with a zero at any point $K$ of Brillouin zone. For the sake of  argument we will assume that $\psi_K$ has one simple zero, but this could be easily generalized to the case of multiple zeros.

Let us notice that at general $K$ there can be only one holomorphic section $\psi_K$. Indeed, if there are two holomorphic linearly independent sections $\psi^{1,2}_K$, their Wronskian must be a holomorphic non-zero double periodic function with specific boundary conditions. 
But this implies that $2K=0 \mod \Lambda$ and the Wronskian is constant. We come to the conclusion  if the other solution exists, it must be meromorphic. 
	
Let us spell out the motivation why a singular wave function satisfying equations \eqref{fundeq} and \eqref{bndc} actually exists. The holomorphic section $\psi_K$ (flat-band wave function) of the bundle $\mathbb{C}^2_K$ forms a subbundle, that we denote as $\gamma$. Then one can consider an exact short sequence:
\begin{gather} \label{esseq}
0 \to \gamma \to \mathbb{C}^2_K \to (\gamma)^\perp \to 0
\end{gather}
where $(\gamma)^\perp = \mathbb{C}^2_K/\gamma$. Roughly speaking we split out the two-dimensional Hilbert space into two one-dimensional ones. The first one is defined to be along the flat-band function $\psi_K$ at each point of the TBG. The second one is chosen to be alongside any other linearly independent wave function. For the sake of argument, one can think of the orthogonal wave function $\psi_\perp$.

The bundles $\gamma$ and $(\gamma)^\perp$ or wave functions $\psi$ and $\psi_\perp$ are one-dimensional wave functions on a periodic TBG and therefore could be assigned (first-) Chern numbers $c_1$. These numbers could be computed in a similar fashion as in the case of the usual Chern numbers in the topological insulators, but where the computations are performed in real space rather than in a momentum space. From mathematical point of view \cite{griffiths1978principles} the Chern number $c_1$ is just the number of zeros minus the number of poles. Since $\mathbb{C}_K$ bundle in some sense trivial, the Chern numbers for the subbundles $\gamma$ and $(\gamma)^\perp$ must satisfy the following relation
 	\begin{gather} 
 		c_1((\gamma)^\perp) = - c_1\left( \gamma \right), \label{crelation}
 	\end{gather}
	suggesting that if $\gamma$ has a holomorphic section, $(\gamma)^\perp$ has a section, but instead of a zero it has a pole. This reasoning is not enough for the proof of its existence, because \eqref{esseq} may not split.  In other words, the wave function $\psi_\perp$ is ill-defined: going around a torus cycle will not only produce a phase, but also add a multiple of $\psi$. From physical point of view it means that we can not simply represent this 2d system as a stack of two topological materials with opposite Chern numbers. 
	
	Luckily, the theory of vector bundles over the Riemann surfaces was actively studied by Donaldson \cite{Donaldson}.  One can show that the short sequence \eqref{esseq} is split over the torus. Below we present a physical construction that can be used to find the solution explicitly. In Appendix \ref{sec:splitting} we prove the existence using algebro-geometric methods.
	
Let us start from a holomorphic section $\psi_K$, that is a solution of the equation \eqref{tbgEq}, satisfies boundary conditions \eqref{bndc}, and has a zero at some point $z_0$. Let us assume for a moment that we somehow managed to find another solution $\phi_K$, that is linearly independent from $\psi_K$. If such a solution exists, the Wronskian $\widetilde{I}(\psi_K, \phi_K)$ should be a meromorphic function
	\begin{gather}
	\det(\psi_K, \phi_K) = \widetilde{I}(z),
	\end{gather}
	that satisfies double-periodic boundary conditions $\widetilde{I}(z+a_{1,2}) = e^{2 i \vec{K} \vec{a}_{1,2}} \widetilde{I}(z)$ (see \eqref{bndc}). Unlike the previous Section, $\phi_K$ might have poles, so we can not conclude that $\widetilde{I}$ is constant. However, an analytic function with this properties exists and is unique up to a normalization factor \cite{griffiths1978principles}. Namely, this function is represented as
	\begin{gather*}
	 \label{FinWron}
	\widetilde{I}(z) = e^{2 i \vec{K} \vec{a}_1 z} \frac{\vartheta\left(z - z_0 ;\tau\right)}{\vartheta(z-z_\infty;\tau)},\,\, z_0-z_\infty = \vec{K}\frac{ \vec{a_2}-\tau \vec{a}_1}{\pi}
	\end{gather*}
	where $\vartheta(z;\tau)$ is a Jacobi theta function. 
	Since $\psi_K$ is finite everywhere, $\phi_K$ must have a pole at the point $z=z_\infty$. From this we get a simple linear equation that $\phi_K$ should satisfy
	\begin{gather}\label{WronksianEq}
	\phi^1_K \psi^2_K - \phi^2_K \psi^1_K = \widetilde{I}(z)
	\end{gather}
	
	Having determined $\widetilde{I}$ let us now construct $\phi_K$.
	At any point $z\in \mathbb{C}/\Lambda$ this equation has at least one solution. Since at point $z=z_0$ both sides of the equation \eqref{WronksianEq} has a simple zero we can analytically continue the solution at this point. Let us pick an arbitrary solution to the eq. (\ref{WronksianEq}) and denote it as $\zeta_K(r)$. Any other solution of the eq. (\ref{WronksianEq}) is
	\begin{gather}\label{lamshift}
	\zeta^\lambda_K(r) = \zeta_K(r) + \lambda(r) \psi_K,
	\end{gather}
	where $\lambda(r)$ is an arbitrary function.
	
	We can derive a relation for the function $\zeta_K(r)$. Namely, we apply an operator $\bar{\partial}$ to the Wronskian to get
	\begin{gather}
	\bar{\partial} \widetilde{I}(z) = \bar{\partial}\det(\psi_K,\zeta_K ) = \notag\\
	= \det(\mathcal{D}\psi_K,\zeta_K) + \det(\psi_K, \mathcal{D}\zeta_K) = \notag\\
	= \det(\psi_K, \mathcal{D}\zeta_K) = 0
	\end{gather}
	This means that in general  $\mathcal{D}\zeta_K$ is proportional to the wave function $\psi_K$
	\begin{gather} \label{SecondEq}
	\mathcal{D}\zeta_K = \eta(r) \psi_K,
	\end{gather}
	for some function $\eta(r)$ which may have singularities.
	To clarify what we have done, the solution $\zeta_K$ is just an arbitrary solution to the equation (\ref{WronksianEq}), and does not satisfy the equation \eqref{tbgEq}. In eq. (\ref{WronksianEq}) we can arbitrary choose $\zeta^1_K$. It could have some singularities. To avoid this problem we set $\zeta^1_K=1$ on the torus. Then the singularities of $\zeta^2_K$ come only from the function $\widetilde{I}(z)$.

	As we discussed before the function $\zeta_K$ is not unique, so we can consider $\zeta^\lambda_K$ from the eq. \eqref{lamshift}. This freedom allows to set the right hand side of the eq. (\ref{SecondEq}) to zero. Indeed,
	\begin{gather}
	\zeta^\lambda_K = \zeta_K + \lambda(r)\psi_K\\
	\mathcal{D}\zeta^\lambda_K(r) = \mathcal{D}\zeta_K(r) + \bar{\partial}\lambda \psi_K(r)=\left[\eta(r)+\bar{\partial}\lambda(r)\right] \psi_K,\notag
	\end{gather}
	Therefore we just need to solve the following equation on a torus
	\begin{gather} \label{eqforlambda}
	\bar{\partial} \lambda = -\eta + C \delta^{(2)}(z-z_0),
	\end{gather}
	with periodic boundary conditions $\lambda(r+a_{1,2}) = \lambda(r)$. The term proportional to $\delta$ function is allowed since $\psi_K$ has a zero at point $z=z_0$. To solve \eqref{eqforlambda} we make a two-dimensional Fourier transform over the torus
	\begin{gather}
		\lambda(k) = \int d^2 \vec{r} \lambda(\vec{r})e^{i k_x x + i k_y y}
	\end{gather}
	Then the equation \eqref{eqforlambda} could be casted as
	\begin{gather}
	\bar{k} \lambda(k) =- \eta(k) + C e^{i k z_0},\quad k = k_x+ i k_y,
	\end{gather}
	This equation has a solution for any $k$ if we tune $C=\eta(0)$. This way
the right hand side is zero at $k=0$, so dividing by $\bar{k}$ we get
	\begin{gather}
	\lambda(k) = - \frac{k \eta(k)}{\left|k\right|^2}, \quad \lambda(0)= 0
	\end{gather}
	Therefore we managed to find a second solution to the equation \eqref{tbgEq} with boundary conditions \eqref{bndc}, that is singular but linearly independent from the holomorphic solution.
	
	One can check that $\lambda(r) \psi_K$ is finite everywhere and therefore the pole of $\zeta_K$ could not be 
removed. We have checked numerically that if one follows the above procedure the resulting wave function has a simple pole and satisfies the system of equations \eqref{tbgEq} and \eqref{bndc}.

	{\it Hidden Landau Levels}. In this section we use units such that the fundamental magnetic flux $\Phi_0 = \frac{h}{e}=1$.
	We have two solutions at the Brillouin point $K$: $\psi^0_K$ with a
zero at a point $z_0$ and $\psi^\infty_K$ with a pole at a point $z_\infty$. We wish to change the basis to these
functions because the original operator $\mathcal{D}$ in \eqref{fundeq} would look very simple in this basis. 
Unfortunately, we cannot do this with original $\psi_K^\infty, \psi_K^0$ because they have a pole and a zero.
We can introduce finite everywhere wave functions $\hat{\psi}^\infty, \hat{\psi}^0$
	\begin{gather}
	\hat{\psi}^\infty = e^{i \vec{K} \vec{a}_1 z -\frac12 B_1 z\bar{z}} \vartheta(z-z_\infty;\tau) \psi^\infty_K,\notag\\ \hat{\psi}^0 = \frac{e^{i\vec{K}\vec{a}_1 z +\frac12 B_1 z\bar{z}}}{\vartheta(z-z_0;\tau)} \psi^0_K,
	\label{hatted_functions}
	\end{gather}
	where $B_1$ is a constant magnetic field corresponding to flux 1 in the moire lattice. Jacobi theta-function
cancels corresponding zero and pole.

	 One can introduce the matrix $S$ which changes the basis:
	\begin{gather}
	S = \left(\hat{\psi}^0, \hat{\psi}^\infty\right) \notag\\
	\det S = \det(\hat{\psi}^0, \hat{\psi}^\infty) = 1 \label{Smatrix}
	\end{gather}
	Then  since $\det S = 1$ we can invert this matrix at each point of the lattice $\mathbb{C}/\Lambda$. This matrix allows to rewrite the Dirac operator as
	\begin{gather}
	\label{HatDirac}
	\hat{\mathcal{D}} = S^{-1} \mathcal{D} S = \begin{pmatrix}
	\bar{\partial} - \frac12 B_1 z & 0\\
	0 & \bar{\partial} +\frac12 B_1 z
	\end{pmatrix},
	\end{gather}
	With the use of transformation $S$ we managed to remove of the
potential $\bar{A}$ from the original Dirac operator $\mathcal{D}$ defined in
the equation \eqref{fundeq} but at the cost of introducing two effective magnetic fields. We would like to point
that the same consideration could be repeated for a holomorphic part of the
Hamiltonian $\mathcal{D}^*$ with the same type of arguments and results.
	
	This shows that in this basis we have just effectively split TBG into
two sheets with effective magnetic field $B_1$. The magnitude of this field is the
same in both sheets, but differs in sign. The form of the equations is exactly the same as for Landau level problem on a torus (see Appendix \ref{sec:landau} and \cite{PhysRevB.31.2529} for the detailed discussion). Since the matrix $S$ is non-singular, the physical solutions for this auxiliary problem must be finite too. In one layer
the effective magnetic field support a wave function with a zero, while in the
other the solution has a pole
and therefore unphysical.  An analogous conclusion was derived
from the different arguments in ref. \cite{ChinesePid}, but in
our case we managed to show
that our system does split into a sum of two systems with non-zero Chern
numbers for a generic potential. 
	
What is the advantage of representation (\ref{HatDirac})?
The key feature of this representation is to allows us to easily
study the system in a external magnetic field.
Such external field corresponds to adding an identity matrix to the anti-holomorphic connection in the eq. \eqref{tbgEq}. It will not be sensitive to the transformation in the eq. \eqref{HatDirac}. Therefore the equation for the zero mode has the following form
	\begin{gather}
	\hat{D}_B f = \begin{pmatrix}
	\bar{\partial} - \frac12 B_1 z + \bar{A}_{U(1)} & 0\\
	0 & \bar{\partial} +\frac12 B_1 z + \bar{A}_{U(1)}
	\end{pmatrix} f = 0, \nonumber \\
	 f = (f_-,f_+)
	\label{HatMagDirac}
	\end{gather}
	where $\bar{A}_{U(1)}$ is a gauge potential for the external magnetic
field that creates magnetic field in the direction perpendicular to the plane
of TBG. Again, physical solutions of these equations are the one with no singularities. We see that we got a simple Landau problem again!
We dedicate Appendix
\ref{sec:landau} for a detailed description of this well-known problem. 

The most important consequence of this, is the emergence of 
extra flat bands.
For simplicity we assume  that $\bar{A}_{U(1)}$ has a flux $\Phi_{ext} = \int
d^2 x \left[\partial \bar{A} - \bar{\partial} A \right] \in \mathbb{Z}$ through
the moire super-lattice. The equation \eqref{HatMagDirac} shows us that the
system decouples into two non-interacting layers with fluxes $\Phi_{tot} \equiv \Phi = \Phi_{ext} \pm 1$.
The result of Appendix \ref{sec:landau} is that for $\Phi_{ext} \ge 1$ there are no flat bands, for $\Phi_{ext}=0$ there
is exactly one, for $\Phi_{ext} \le -1$ there are $2|\Phi_{ext}|$ flat bands.

	Note that we have studied only the anti-holomorphic part of the
Hamiltonian (\ref{tbgHam}). The holomorphic part(the other chirality) exhibits
the same properties but for $\Phi_{ext} \rightarrow -\Phi_{ext}$. It means that in total
there are $2\left|\Phi_{ext}\right|$ flat bands for $\left|\Phi_{ext}\right| > 0$ and for
$\Phi_{ext}=0$ there are only 2 flat bands. We would like to point that if we did not take into account the hidden wave function we would expect to have $\left|\Phi_{ext}\right| +2$ flat bands in the presence of the external magnetic field.

	{\it Physical consequences and conclusion.}
	Let us summarize our key findings. We started from CS-CM-type
	Hamiltonian (\ref{tbgHam}) with a \textit{generic} $\bar{A}$, and assumed that it has a flat band. We proceeded by deriving an
	extra (nonphysical) zero-mode $\psi^\infty_K$ of the Hamiltonian
	(\ref{tbgHam}). This solution let us define the transformation $S$(eq. (\ref{Smatrix})) and represent the original
	Dirac operator in the form (\ref{HatDirac}). This new form is very simple
	and it allowed us to explicitly demonstrate the emergence of extra flat bands in the presence of external transverse magnetic field. 
	
	The TBG is believed to be approximately described by the CS-CM model\cite{Bistritzer2010,Bistritzer2011}. As it was shown in the paper \cite{Grisha} such system posses a flat band solution at $\theta=1.05^\circ$. Since we derived some general properties of the solutions of CS-CM model we can argue that our results are applicable to the TBG.
	At the first magic angle $\theta=1.05^\circ$, unity fundamental flux through moire lattice correspond to magnetic field of about $28$ T. Such magnetic fields are accessible, hence our prediction of extra flat bands can be, in principle, verified on a experiment.

{\it Acknowledgments} 
FKP dedicates this paper to the memory of his father, Popov Kalina Fedorovich. We would like to thank Grisha Tarnopolsky and Igor R. Klebanov for the collaboration at the early stages of the work. We would like to thank Vladyslav Kozii, Anastasia Aristova, Alexander Avdoshkin, Nikita Sopenko, Dumitru Calugaru for useful comments and discussions. We would like to thank Alexander Gorsky, Emil Akhmedov and Peter Czajka for useful comments on the draft of the paper. AM also would like to thank C.~King for moral support.
The research of FP was supported in part by the US NSF under Grants No. PHY-1620059 and PHY-1914860.
The work of AM was supported by the Air Force Office of Scientific Research under award number FA9550-19-1-0360.   It  was  also  supported  in  part  by  funds  from  the  University  of California.   

\appendix   
 
   \setcounter{equation}{0}	
	\renewcommand{\Im}{\operatorname{Im}}
	\section{Landau Levels on torus.} 
\label{sec:landau}

	In this section we briefly review the wave functions on torus, we will mostly follow Haldane and Rezayi \cite{PhysRevB.31.2529}.
	We consider a complex torus $\mathbb{C}/\Lambda, \Lambda=\left\{n+m\tau; n,m\in\mathbb{Z}\right\}$, and want to find solutions to the following equation
	\begin{gather}
	\label{D_B_eq}
	\hat{D}_B f = \left(\bar{\partial} + \frac12 e B z\right) f = 0 , \quad F = \bar{\partial} A - \partial \bar{A} = B.
	\end{gather} 
	which is either of the two equations in \eqref{HatMagDirac}.
	To establish the boundary conditions we consider a shift of $z$ by a lattice vector $a_i=1,\tau$ to get
	\begin{gather}
	\left(\bar{\partial} + \frac12 e B z + \frac12 e B a_i \right) f = 0 
	\end{gather}
	To remove the change in the gauge potential, we should make a gauge transformation 
	\begin{gather}
	f \to f e^{-\frac12 e B a_i \bar{z} + \frac12 e B \bar{a}_i z},
	\end{gather}
	this can be used to define the boundary conditions. Namely,
	\begin{gather}
	    T_1: f(z+1)= \psi(z) e^{-\frac12 e B \bar{z} + \frac12 e B z}, \notag\\
	     T_\tau: f(z+\tau)=  f(z) e^{-\frac12 e B \tau\bar{z} + \frac12 e B \bar{\tau} z}    \label{bndc3}
	\end{gather}
	We should check the consistency of this boundary conditions, that $T_1 T_\tau = T_\tau T_1$. One can check that 
	\begin{gather}
	    T_1 T_\tau f(z) = e^{e B \left(\tau - \bar{\tau}\right)} T_\tau T_1 f(z)
	\end{gather}
	The difference between this phases is $e^{e B \left(\tau - \bar{\tau}\right)}=1$. That gives a condition for the consistent boundary conditions \eqref{bndc3}
	\begin{gather}
	\pi \Phi = e B \Im \tau, \quad eB = \frac{\pi \Phi}{\Im \tau}, \quad \Phi \in \mathbb{Z}
	\end{gather}
	This conditions give that the integral over a fundamental period is equal to $\Phi = \frac{1}{2\pi} \int F d^2 z =  \frac{eB}{\pi} \Im \tau$. 
	
	Then if we can consider a general $k$ from Brillouin zone to get
	\begin{gather}
	\left(\bar{\partial} + \frac12  e B z\right)f_k(z) = 0, \quad \notag\\
	f_k(z + a_i) =  f_k(z) e^{- \frac12 e B a_i \bar{z} +\frac12 e B \bar{a}_i z + i (k,a_i)} \label{lantor}
	\end{gather}
	Where $(k,z) = k_x x + k_y y = \operatorname{Im} k \bar{z}$, $k = k_x + i k_y$, $z= x + i y$.

	For $\Phi=-1$ the equation \eqref{lantor} is easy to solve, we get
	\begin{gather*}
	f_{k,-1}(z) = \vartheta\left(z + \frac{i k}{2 e B}; \tau\right) e^{\frac12 e B(z + \frac{i k }{2 e B})^2  + i \frac12 \bar{k} z -\frac12 e B z \bar{z}}
	\end{gather*}

	The zero of this function is located at
	\begin{gather}
	z_0= \frac12 + \frac12 \tau - \frac{i k}{2 e B} 
	\end{gather}
	
Whereas for $\Phi=+1$ we get a solution
	with a pole:
	\begin{gather*}
	 f_{k,+1}(z) = \frac{1}{\vartheta\left(z + \frac{i k}{2 e B}; \tau\right)} e^{\frac12 e B(z + \frac{i k }{2 e B})^2+i \frac12 \bar{k} z -\frac12 e B z \bar{z}}
	\end{gather*}
	
	If $\Phi \neq -1$, the solution is just 
	\begin{gather}
	f_{K,\Phi}(z) = \prod\limits_{k_1+\ldots + k_\Phi = K}f_{k_i,-1}(z) \label{lanPhi}
	\end{gather}
	It seems that there is now an infinite number of the wave functions at given $K$. One can show that there is only a finite number of the linearly independent solutions \eqref{lanPhi}.
	
	To compute this dimension we can use the Riemann--Roch formula \cite{griffiths1978principles} for the operator $\hat{D}_B$ in eq. (\ref{D_B_eq}). This gives that $\dim \ker \hat{D}_B = \Phi$, so when $\Phi$ is negative we don't have any finite solutions for the Landau levels at any point of the Brillouin zone. The case $\Phi=0$ is special - there is some zero modes but only at special points of the Brillouin zone.
	
	\section{ The splitting of short exact sequence and \v{C}ech cohomology.}
\label{sec:splitting}
	In this subsection we would like to clarify the existence of the second solution from the cohomology point of view. 
	Although this approach is a bit involved, it is mathematically rigorous and could be generalized to higher genus Riemann surfaces \cite{Gunning}.
	
	We start with the rigorous formulation of the problem. Assume that we have a Riemann surface $\mathcal{M}$ with some holomorphic vector bundle $\pi$ of rank 2, $\pi: E \to \mathcal{M}$ and some connection $\bar{A}$. Namely, we have a covering of the Riemann surface with open subsets $\left\{U_\alpha \right\}$ where the vector bundle could be trivialized
	\begin{gather}
	\mathcal{M} = \bigcup\limits_\alpha\, U_\alpha,\quad \left.E\right|_{U_\alpha} \approx U_\alpha \times \mathbb{C}^2.
	\end{gather}
	When we move from one covering $U_\alpha$ to another $U_\beta$, we need to glue the section with holomorphic gluing functions $g^0_{\alpha\beta}$, $\bar{\partial}g^0_{\alpha\beta}=0$. The connection $\bar{A}_\beta$ transforms as
	\begin{gather}
	\bar{A}_\beta = \hat{g}^{-1}_{\alpha\beta} \bar{A}_{\alpha} \hat{g}_{\alpha\beta}+\hat{g}^{-1}_{\alpha\beta} \bar{\partial} \hat{g}_{\alpha\beta}
	\end{gather}
	We want to find holomorphic sections of these vector bundles: a collection of functions $\left\{\psi_\alpha\right\}$, such that the following conditions are satisfied
	\begin{gather}
	\quad \psi^0_\alpha = g^0_{\alpha\beta} \psi^0_\beta,\quad \left(\bar{\partial}+\bar{A}_\alpha\right) \psi^0_\alpha = 0 \label{coneq}
	\end{gather}
	We can get rid of the connection $\bar{A}_\alpha$ by solving the equation \eqref{coneq} at each covering and performing the gauge transformation.

    Then we can generally study the following problem
	\begin{gather}\label{redefbundle}
		\psi_\alpha = g_{\alpha\beta} \psi_\beta, \quad \bar{\partial} \psi_\alpha=0.
	\end{gather}
	So we just need to find a {\it meromorphic} sections of the vector bundle $E$ defined by cocycles $g_{\beta\alpha}$ in the assumption that we have a {\it holomorphic} sections of the bundle \eqref{redefbundle} $\psi^h$. Namely, we have a collection of holomorphic functions $\psi_\alpha^h(z)$ defined at each coverings and satisfy boundary conditions
	\begin{gather}
	\psi_\alpha^h = g_{\alpha\beta} \psi_\beta^h
	\end{gather}
	We want this functions $\psi_\alpha^h$ to be non-zero at any coverings of $\mathcal{M}$. Whenever we encounter a zero in some covering $U_\alpha$, $\psi^h_\alpha(z^\alpha_0)=0$, we redefine holomorphic section and gluing functions $\hat{\psi}_\alpha^h =\frac{1}{z-z^\alpha_0} \psi_\alpha^h$ and $\hat{g}_{\alpha\beta} = \gamma^1_{\alpha\beta} g_{\alpha\beta}, \gamma^1_{\alpha\beta} = \frac{z - z^\alpha_0}{z-z^\beta_0}$ . This new function is nowhere zero and changes as
	\begin{gather}
	\hat{\psi}_\alpha^h = \gamma^1_{\alpha\beta} g_{\alpha\beta} \hat{\psi}_\beta^0, \quad\text{where}\quad \gamma^1_{\alpha\beta} \in \mathbb{C},\quad \gamma^1_{\alpha\beta}\gamma^1_{\beta\gamma}\gamma^1_{\gamma\alpha}=1
	\end{gather}
	Since this section is nowhere zero we can find another set of holomorphic functions that is linearly independent from $\hat{\psi}_\alpha$ at each point. We call this set of functions as $\hat{\psi}^\infty_\alpha$ and with analgous procedure introduce $\gamma^2_{\alpha\beta}$ to remove all zeros it can possible have. 
	Because of this at each covering we can change basis to $\hat{\psi}_\alpha$ and $\hat{\psi}^\infty_\alpha$. One can check that gluing functions in this new basis of the vector bundle become
	\begin{gather}	
	\hat{g}_{\alpha\beta} = \begin{pmatrix}
	\gamma^2_{\alpha\beta} & h_{\alpha\beta}\\
	0 & \gamma^1_{\alpha\beta}
	\end{pmatrix}, \quad \text{where} \notag\\ \gamma^2_{\alpha\beta}\gamma^2_{\beta\gamma}\gamma^2_{\gamma\alpha}=1 \quad \text{and} \quad h_{\alpha\gamma} = \gamma^2_{\alpha\beta}h_{\beta\gamma} + h_{\alpha\beta} \gamma^1_{\beta\gamma}
	\end{gather}
	If we got that $h_{\alpha\beta}=0$ then the function $\hat{\psi}^\infty_\alpha$ would change through each other as
	\begin{gather}
	\hat{\psi}^\infty_\alpha = \gamma^2_{\alpha\beta} g_{\alpha\beta}\hat{\psi}^\infty_\beta,
	\end{gather}
	and define a legimate section of the vector bundle $E$.
	Since the set of function $\gamma^2_{\alpha\beta}$ represents a line bundle, it has a meromorphic section: a set of meromorphic functions $f_\alpha(z)$ with property $\gamma^2_{\alpha\beta} = \frac{f_\beta}{f_\alpha}$. Then new functions
	\begin{gather}
	\psi^\infty_\alpha = f_\alpha \hat{\psi}^\infty_\alpha
	\end{gather}
	are holomorphic everywhere and transforms as 
	\begin{gather}
	\psi^\infty_\alpha = g_{\alpha\beta} \psi^\infty_\beta,
	\end{gather}
	And therefore represents a legitimate section of the original vector bundle $E$, but contains a pole at some point.

	Let us show that we can get rid of $h_{\alpha\beta}$ by a proper redefinition of the arbitrary chosen $\hat{\psi}^\infty_\alpha$. Namely, we notice that the choice of $\hat{\psi}^\infty_\alpha$ is not unique. At each covering we can make a change
	\begin{gather}
	\hat{\psi}^\infty_\alpha \to \hat{\psi}^\infty_\alpha + h_\alpha(z) \hat{\psi}^0_\alpha
	\end{gather}
	It changes gluing functions as
	\begin{gather} \label{cohchan}
	h_{\alpha\beta} \to h_{\alpha\beta} + \gamma^2_{\alpha\beta}h_\beta(z) - h_\alpha(z)\gamma^1_{\alpha\beta}
	\end{gather}
	This gives that $h_{\alpha\beta}$ belongs to $H^1(\mathcal{O}(\gamma_1\gamma^{-1}_2))$ and by Serre duality are dual to $H^0(\mathcal{O}(\kappa\gamma_1^{-1}\gamma_2))$, where $\kappa$ is a tangent line bundle. This line bundle does not have any holomorphic section if its Chern class is negative. We get
	\begin{gather}
	\label{H0_calc}
	c_1(\kappa \gamma_1^{-1} \gamma_2) = 2 g - 2 - c(\gamma_1) + c(\gamma_2) =\notag\\  
	= 2 g - 2 - 2 c(\gamma_1) = -2 c(\gamma_1) < 0
	\end{gather}
	and we used $g=1$(torus), $c_1(\gamma_1) \ge 1$ ($\hat{\psi}_\alpha$ has at least one simple zero) and $c_1(\gamma_1) + c_1(\gamma_2)=0$(
	consequence of eq. (\ref{crelation})).
	
	Since $H^1(\mathcal{O}(\gamma_1\gamma^{-1}_2))=0$ the cohomology class represented by $h_{\alpha\beta}$ is trivial. Meaning, that we can always pick $h_\alpha$ such that $h_{\alpha\beta} = 0 $ in the eq. \eqref{cohchan}. Then as we discussed above $\hat{\psi}^\infty_\alpha$ will represent a meromorphic section of the vector bundle $E$. This procedure could be generalized to higher genus Riemann surfaces and other vector bundles.

\end{document}